\documentclass[prl,12pt,tightenlines,showpacs,showkeys]{revtex4}
\usepackage{latexsym}
\usepackage{graphicx}
\usepackage{amsmath}
\usepackage{amsbsy}
\usepackage{amsthm}
\usepackage{bbm}
\usepackage{bm}

\newtheorem{lem}{Lemma}

\begin{document}

\title{Bound entanglement in the set of Bell state mixtures\\ of two qutrits}
\author{Reinhold A. Bertlmann} \email{reinhold.bertlmann@univie.ac.at}
\author{Philipp Krammer} \email{philipp.krammer@univie.ac.at}
\affiliation{Faculty of Physics, University of Vienna,
Boltzmanngasse 5, A-1090 Vienna, Austria}

\begin{abstract}

We investigate the entanglement properties of a three--parameter
family of states that are part of the \emph{magic simplex} of two
qutrits, which is a simplex of states that are mixtures of maximally
entangled two--qutrit \emph{Bell states}. Using entanglement witnesses
we reveal large regions of bound entangled and separable states.

\end{abstract}

\keywords{entanglement, bound entanglement, entanglement witness, qutrit}
\pacs{03.67.Mn, 03.65.Ca, 03.65.Ta, 03.67.Hk}

\maketitle

Entangled states are the basic ingredients of quantum information
and quantum communication. For pure states of any dimension and any
number of particles it is easy to determine if a state is entangled
or not, the property is revealed by its reduced density matrices
(see, e.g., Ref.~\cite{bruss02}). For mixed states, however, it is
in general difficult to decide whether a given density matrix is
entangled or separable. Since quantum mechanical states are in
practice rather mixed than pure, one seeks for criteria to detect
entanglement and separability. Additionally, entangled states can be
classified according to their distillability -- a state is called
\emph{distillable} if from many copies of the same not maximally
entangled state we can distill fewer maximally entangled states with
statistical local operations and classical communication (SLOCC).
Interestingly, not all entangled states are distillable. For states
of higher dimension than $2 \times 2$ and $2 \times 3$, there exist
states that are entangled but not distillable, so-called bound
entangled states \cite{horodecki97a, horodecki98, horodecki99c,
rains99, baumgartner07, horodecki07}. Apart from being a
mathematical oddity bound entangled states physically characterize a
set of states that are an obstacle for quantum information
processing tasks, since once a system evolved into a bound entangled
state, there is no way to retain a maximally entangled state with
SLOCC. Nevertheless, concerning their usefulness, bound entangled
states are not equivalent to separable states since their
entanglement can be ``activated'' when used together with
distillable entangled states \cite{horodecki99c}. It is the
intention of this paper to present a certain family of two--qutrit
states that lie within the \emph{magic simplex}
\cite{baumgartner07a} -- this family reveals a high geometric
symmetry, can be classified according to its entanglement properties
and contains large regions of bound entanglement.

We consider a Hilbert-Schmidt space ${\cal A}_A \otimes {\cal A}_B$
of operators on the finite dimensional bipartite Hilbert space ${\cal
H}_A \otimes {\cal H}_B$, with dimension $d_A \times d_B$, $D := d_A
d_B$. States $\rho$ (i.e. density matrices) are
elements of ${\cal A}_A \otimes {\cal A}_B$ with the properties
$\rho^\dag = \rho$, Tr $\rho = 1$ and $\rho \geq 0$. A scalar product on
${\cal A}_A \otimes {\cal A}_B$ is defined by $\left\langle A,B \right\rangle
= \textnormal{Tr}\, A^\dag B$ with $A,B \in {\cal A}_A \otimes {\cal A}_B$.

Two basic concepts in the context of entanglement detection are
entanglement witnesses \cite{horodecki96, terhal00, terhal02,
bertlmann02, bertlmann08} and the PPT Criterion (positive partial
transposition) \cite{peres96, horodecki96}. An entanglement witness
$A$ detects the entanglement of states due to the convexity of the
set of separable states $S$: A state $\rho$ is entangled iff there
exists a Hermitian operator $A$ such that
\begin{eqnarray}
    \left\langle \rho,A \right\rangle \;=\; \textnormal{Tr}\, \rho A
    & \;<\; & 0 \,, \label{defentwitent} \\
    \left\langle \sigma,A \right\rangle = \textnormal{Tr}\, \sigma A & \;\geq\; & 0 \qquad
    \forall \,\sigma \in S \,. \label{defentwitsep}
\end{eqnarray}
For a given state $\rho$ it is easy to find operators that provide
Eq.~\eqref{defentwitent}, but in general difficult to show
Eq.~\eqref{defentwitsep}. Nevertheless it proves useful in many
cases. If there exists a separable state $\tilde{\sigma}$ for which
${\rm Tr} \, \tilde{\sigma} A = 0$, then $A$ is called \emph{optimal
entanglement witness}. Optimal entanglement witnesses correspond to
hyperplanes given by ${\rm Tr} \, \tilde{\sigma} A = 0$ which are
tangent to the set of separable states $S$.

An operational criterion that is easy to apply is the PPT Criterion:
A separable state $\sigma$ stays positive under partial
transposition,
    $
        \sigma^{T_B} := \left( \mathbbm{1} \otimes T \right) \sigma \,\geq\, 0 \,.
    $
We call a state PPT that is positive under partial transposition,
and NPT a state that is not. Note that the PPT criterion is a
necessary criterion for separability \cite{peres96}, that means if a
state $\rho$ is NPT, it has to be entangled. But if it is PPT, this
does not automatically imply that it is separable, this is true for
dimensions $2 \times 2$ and $2 \times 3$ only \cite{horodecki96}.
Amazingly, entangled PPT states are not distillable, thus
\emph{bound entangled} \cite{horodecki98, rains99, horodecki07}.
Clearly the PPT Criterion cannot distinguish PPT entangled from
separable states, therefore other entanglement criteria have to be
used, e.g. the entanglement witness criterion, which is the method
that we will apply in this paper.

In Ref.~\cite{bertlmann08} we presented a method to detect bound
entanglement that is based on a geometrical approach: Consider a
line of states $\rho_\lambda$ inside the convex set of the PPT
states between a PPT state $\rho_{PPT}$ and the maximally mixed
state,
\begin{equation} \label{rholambda}
    \rho_\lambda \,:=\, \lambda \,\rho_{\rm{PPT}} +  \frac{(1-\lambda)}{D}
    \mathbbm{1}_{\rm{D}} \,, \quad 0 \leq \lambda \leq 1 \,,
\end{equation}
and an operator
\begin{equation} \label{clambda}
C_\lambda \,=\, \rho_\lambda - \rho_{\rm{PPT}} - \langle \rho_\lambda ,
\rho_\lambda -
    \rho_{\rm{PPT}} \rangle \mathbbm{1}_{\rm{D}} \,.
\end{equation}
The operator $C_\lambda$ is constructed geometrically in such way
that we have Tr $C_\lambda \rho_\lambda = 0$ and Tr $C_\lambda
\rho_{\rm{PPT}} < 0$ (here $\lambda < 1$, the limit $\lambda
\rightarrow 1$ is discussed later on). It divides the whole state
space into states $\rho_n$ for which Tr $C_\lambda \rho_n < 0$ and
states $\rho_p$ for which Tr $C_\lambda \rho_p \geq 0$, see
Ref.~\cite{bertlmann05}. If $C_\lambda$ is an entanglement witness
then all states $\rho_n$ on one side of the plane are entangled. For
a particular entanglement witness $C_{\tilde{\lambda}}$ all states
$\rho_\lambda$ are PPT and entangled for $\tilde{\lambda} < \lambda
\leq 1$, thus bound entangled. Clearly the difficulty of the
presented method lies in proving that $C_\lambda \geq 0 \ \forall \
\sigma \in S$. Different lines  \eqref{rholambda} are distinguished
by different ``starting states'' $\rho_{\rm{PPT}}$ which will be
called ``starting points'' throughout this text.

Next we apply the above method to a three--parameter
family of the magic simplex of two qutrits. The magic simplex ${\cal
W}$ of bipartite qutrits (dimension $3 \times 3$) is introduced in
Ref.~\cite{baumgartner06}, extended to qudits (dimension $d \times
d$) in Ref.~\cite{baumgartner07a}, reviewed and discussed in
Ref.~\cite{baumgartner07}. It is defined as the set of all states
that are a mixture of \emph{Bell states} $P_{nm}$,
\begin{equation}
    {\cal W} \,:=\, \left\{ \sum_{n,\,m=0}^{d-1} q_{nm} P_{nm} \ | \ q_{nm}
    \geq 0, \ \sum_{n,\,m} q_{nm}=1 \right\},
\end{equation}
where
\begin{equation} \label{bellstates}
    P_{nm} \,:=\, (U_{nm} \otimes \mathbbm{1}) | \phi^+_d \rangle \langle
    \phi^+_d | (U_{nm}^\dag \otimes \mathbbm{1}) \,.
\end{equation}
The vector state $| \phi^+_d \rangle$ denotes the maximally
entangled state $\left| \phi^+_d \right\rangle = \frac{1}{\sqrt{d}}
\,\sum_j \left| j \right\rangle \otimes \left| j \right\rangle$ and
the \emph{Weyl operators}
$U_{nm} = \sum_{k=0}^{d - 1} e^{\frac{2 \pi i}{d}\,kn} \,| k \rangle
    \langle (k+m) \,\textrm{mod}\,d|$
form an orthogonal basis $\{U_{nm}\}$ $(n,m = 0,1, \ldots ,d - 1)$
of the Hilbert-Schmidt space \cite{bennett93, narnhofer06}.

The indices $n$ and $m$ can be viewed as coordinates in a discrete
phase space, and the points for fixed $n$ or $m$ constitute lines
within. The magic simplex is highly symmetric and exhibits the same
geometry under a line change (for details see
Refs.~\cite{baumgartner06, baumgartner07}). It thus represents the
eight--dimensional analogue of the three--dimensional simplex, a
tetrahedron, of two--qubits which is the set of states that are
mixtures of the four two--qubit Bell states \cite{horodecki96b,
vollbrecht01, bertlmann02, bertlmann07}. The two--qudit Bell states
\eqref{bellstates} play an important role in extensions of quantum
communication procedures to higher dimensional systems -- e.g. in
the quantum teleportation protocol \cite{bennett93}. They constitute
the corresponding maximally entangled orthogonal basis and the Weyl
operators $U_{nm}$ are a generalization of the Pauli operators, in
the sense that Bob has to apply one of the $d^2$ operators
(including the identity operator) to obtain the teleported state.

\begin{figure}
  \includegraphics[width=0.7\textwidth]{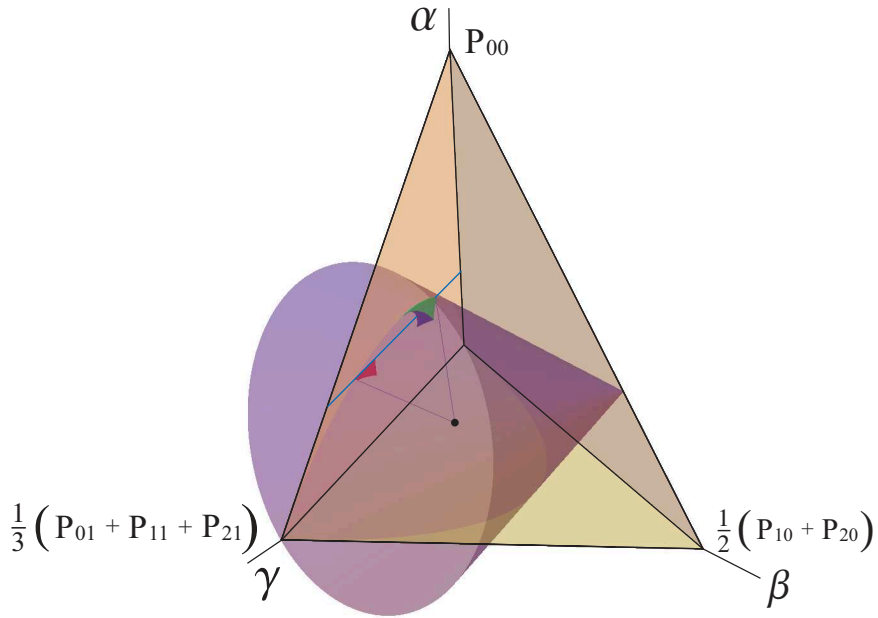}\\
  \caption{(Color online) In the parameter space representing the Hilbert-Schmidt space geometry
    the states $\rho_{\alpha,\beta,\gamma}$
    \eqref{famstates} lie inside a pyramid with triangular base, the PPT points
    form a cone. The intersection region includes the PPT states, i.e. separable
    and bound entangled states. The maximally mixed state is represented by the black dot.
    The line (blue) on the boundary plane indicates the Horodecki states
    \eqref{horstates}, the planar section (red) jutting out of the Horodecki line shows the
    bound entangled states that are detected using starting points on the Horodecki line only.
    The protruding volume (blue/green)
    is the volume of bound entangled states that can be obtained with starting points
    deviating from the Horodecki line. In fact the picture is symmetrical for positive
    and negative values of $\gamma$, but for the sake of clarity only one volume is
    depicted. All parameter axes are chosen non-orthogonal to each other such that they become
    orthogonal to the boundary of the positivity region in order to reproduce the
    symmetry of the magic simplex.} \label{figconepyramid}
\end{figure}

We want to consider a three--parameter family of states, introduced
in Ref.~\cite{bertlmann08}, of the magic simplex that are a mixture
of the maximally mixed state and two phase space lines,
\begin{equation} \label{famstates}
    \rho_{\alpha,\beta,\gamma} \;:=\; \frac{1-\alpha -\beta -\gamma}{9}
    \mathbbm{1} + \alpha P_{00} + \frac{\beta}{2} \left( P_{10} +
    P_{20} \right) + \frac{\gamma}{3} \left( P_{01} +P_{11}+P_{21}
    \right)\,.
\end{equation}
The parameters are constrained by the positivity requirement
$\rho_{\alpha,\beta,\gamma} \geq 0$, which results in $\alpha \leq 7
\beta /2 +1 -\gamma, \ \alpha \leq -\beta +1 -\gamma, \ \alpha \leq
-\beta +1 +2\gamma,$ and $\alpha \geq \beta/8 - 1/8 + \gamma/8$. The
constraints geometrically represent a pyramid with triangular base,
see Fig.~\ref{figconepyramid}. The PPT Criterion selects those
points $(\alpha,\beta,\gamma)$ that correspond to positive operators
when subjected to partial transposition. We obtain the constraints
$\alpha \leq - \beta - 1/2 + \gamma /2$, as well as $\alpha \leq (
-2 + 11\beta - \gamma + 3 \sqrt{\Delta})/16$ and $\alpha \leq ( -2 +
11\beta - \gamma - 3 \sqrt{\Delta})/16$, where we defined $\Delta =
4 + 9 \beta^2 + 4\gamma - 7\gamma^2 - 6\beta (2 + \gamma)$, which
form a cone that intersects the pyramid, and in the intersection
region lie the PPT states, see Fig.~\ref{figconepyramid}.

A parameterized line of states $\rho_{\alpha,\beta,\gamma}$ situated on the
boundary plane $\alpha = 7 \beta /2 +1 -\gamma$ (which we will from now on refer to
as the \emph{boundary plane}) is given by
\begin{equation} \label{horstatessimplex}
    \rho_b \;\equiv\; \rho_{\alpha, \beta, \gamma} \qquad\mbox{with}\quad \alpha =
    \frac{6-b}{21}, \;\beta = -\frac{2b}{21}, \;\gamma = \frac{5-2b}{7} \,,
\end{equation}
which exactly corresponds to the states
\begin{equation} \label{horstates}
    \rho_b \;=\; \frac{2}{7} \left| \phi_+^3 \right\rangle \left\langle
    \phi_+^3 \right| \,+\, \frac{b}{7} \, \sigma_+ \,+\, \frac{5 - b}{7} \,
    \sigma_- \,, \qquad 0 \leq b \leq 5
\end{equation}
that were introduced in Ref.~\cite{horodecki99c}. We call this
one--parameter family \emph{Horodecki states} or \emph{Horodecki
line}. It is shown in Ref.~\cite{horodecki99c} that these states
are entangled for $4 < b \leq 5$, bound entangled for $3 < b \leq 4
$ and separable for $2 \leq b \leq 3$. For convenience we express the parameter
$b$ in terms of $\gamma$ to obtain a parametrization of the Horodecki line with $\gamma$.

In Ref.~\cite{bertlmann08} we discovered planar regions of bound entanglement
that go beyond the Horodecki line with the method discussed before, where the
starting points $\rho_{\rm PPT}$ of the line $\rho_\lambda$ \eqref{rholambda}
are chosen as the bound entangled Horodecki states. The inequality \eqref{defentwitsep}
is proven by expressing the operator $C_\lambda$ and the separable states in
terms of Weyl operator tensor products. In this way we arrive at a sufficient condition
for $C_\lambda$ such that Eq.~\eqref{defentwitsep} is satisfied:
\begin{lem} \label{lemqutrit}
    For any Hermitian operator $C$ of a bipartite Hilbert-Schmidt
    space of dimension $d \times d$ that is of the form
    \begin{equation} \label{lemqutritc}
        C \;=\; a \left( (d-1) \, \mathbbm{1}_{\rm{d^2}} \,+\, \sum_{n,m=0}^{d - 1} c_{nm}
            \,U_{nm} \otimes U_{-nm} \right), \quad a \in \mathbbm{R}^+, \
            c_{nm} \in \mathbbm{C}
    \end{equation}
    the expectation value for all separable states is positive,
    \begin{equation} \label{ewineqc}
        \langle \rho , C \rangle \,\geq\, 0 \quad \forall \, \rho \in S \,,
        \quad \mbox{\textrm{if}} \quad |c_{nm}| \,\leq\, 1 \quad \forall \, n,m \,.
    \end{equation}
\end{lem}

If we choose the starting points not only on the Horodecki line, but
also in a neighborhood of the line on the boundary plane, we can detect
volumes of bound entanglement, see Fig.~\ref{figconepyramid}. These starting points
$\rho_{\rm plane}$ are parameterized with an additional parameter $\epsilon$
that accounts for the amount of deviation from the line,
\begin{equation} \label{planestates}
    \rho_{\rm{plane}} \;\equiv\; \rho_{\alpha, \beta, \gamma} \quad \mbox{with} \quad
    \left(\alpha = \frac{1+\gamma+\epsilon}{6} \,, \beta=
    \frac{-5+7\gamma+\epsilon}{21} \,,\gamma \right), \quad \epsilon \in \mathbbm{R}\,.
\end{equation}
The parameters $\epsilon$ and $\gamma$ of the states $\rho_{\rm
plane}$ are restricted by the condition $|c_{nm}| \leq 1 \ \forall n,m$
of Lemma~\ref{lemqutrit}. The coefficients $c_{nm}(\gamma)$ of $C_\lambda$
fulfil $c_{nm}(\gamma) = c_{nm}^\ast(-\gamma)$, hence Lemma~\ref{lemqutrit}
can be applied for positive and negative values of $\gamma$, and we obtain
a symmetric picture. We restrict ourselves to positive values of $\gamma$,
obtaining the ``negative side'' by applying $\gamma \rightarrow -\gamma$.
The starting points on the plane can
be seen in Fig.~\ref{figconepyramid} as the points where the volumes
of bound entangled states emerge from the boundary plane. Note that
only the states inside the volumes are bound entangled for sure,
states $\rho_\lambda^{\rm surface}$ on the surface can be either separable
or bound entangled. The reason is that for points $\rho_\lambda^{\rm surface}$
we have ${\rm Tr} \,\rho_\lambda^{\rm surface} C_\lambda^{\rm surface} = 0$
and in general we do not know whether the operator $C_\lambda^{\rm surface}$ defines
a tangent plane to the set of separable states, i.e. whether it is already optimal.
The detected regions of bound entanglement, the ``old'' planar region and the
``new'' volumes, are plotted in Fig.~\ref{figconepyramid}.

Entanglement witnesses of the form \eqref{clambda} correspond to
planes in our three--dimensional illustration of the Hilbert-Schmidt
space. They detect not only the entanglement of states on lines
$\rho_\lambda$ \eqref{rholambda}, but also of all states lying on the
same side of the plane that includes the starting points $\rho_{\rm PPT}$
of $\rho_\lambda$. Consider for example a starting point $\rho_{\rm plane}^1$
\eqref{planestates} on the boundary plane that represents the lower tip
of the bound entangled volume in Fig.~\ref{figconepyramid},
it is given by $(\epsilon = -1/4, \gamma = 1/4)$.
Using Lemma~\ref{lemqutrit} we find that for $\rho_{\rm plane}^1$ the operator
$C_1 := C_\lambda/\lambda (1-\lambda )$ is an entanglement witness
for $\lambda \rightarrow 1$ and has a simple matrix form which we can
easily calculate from Eq.~\eqref{clambda}.\\

\begin{figure}
  \includegraphics[width=0.25\textwidth]{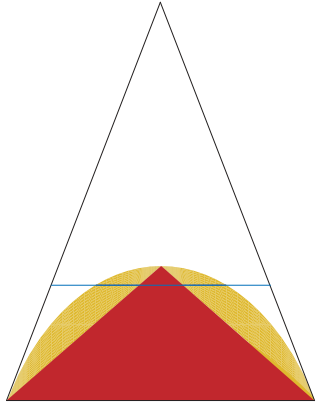}\\
  \caption{(Color online) The entanglement properties of states $\rho_{\alpha,\beta,\gamma}$ on the
  boundary plane $\alpha = (7/2) \beta +1 -\gamma$. The separable states are situated
  in the triangular region (red), the PPT entangled (bound entangled) states
  lie in between the triangular and parabolic region (orange), and the
  remaining states are  NPT entangled. Interestingly, the Horodecki line (blue) runs
  through all entanglement characteristics.} \label{figplanesep}
\end{figure}

Now we explain how to detect further bound entangled states
and the region of separability.
The trace ${\rm Tr} \, C_1 \rho_{\rm plane}^1 = 0$ provides the plane
equation $Pl_1: \, \alpha = 2(1+2 \beta -\gamma )/5$.
Plane $Pl_1$ intersects the boundary plane on the line $l_a: \ \beta = 2(-1+\gamma )/9$,
the cone of PPT states intersects the boundary plane on the curve
$l_b: \beta = (-4+3 \gamma + \sqrt{4-3 \gamma ^2})/9$, and $l_a$ and $l_b$ cross
each other at $\gamma = 0$ and $\gamma = 1$. These crossing points are
separable states: $\gamma = 0$ corresponds to two--parameter states
\cite{baumgartner06, bertlmann08}, in this case it is proved in
Ref.~\cite{baumgartner06} that all PPT states are separable, and for
$\gamma = 1$ we have $\alpha = \beta = 0$, i.e. an equal mixture of the Bell states
$P_{01}$, $P_{11}$, and $P_{21}$, which is separable too \cite{baumgartner06}.
The hyperplane of the entanglement witness cuts the boundary plane at the line $l_a$,
it contains two separable states, therefore it represents an optimal entanglement
witness that has to be tangent to the set of separable states $S$.
Since $S$ is convex the line $l_a$ has to describe the boundary of the
separable states between $\gamma=0$ and $\gamma=1$. Therefore the whole
region between $l_a$ and $l_b$ has to be bound entangled, it is given by
$0 < \gamma < 1$, $\,2(-1+\gamma )/9 < \beta < (-4+3 \gamma +\sqrt{4-3 \gamma ^2}/9)$
and $\alpha = 7\beta /2 +1 -\gamma$. For the Horodecki
states we thus find bound entanglement for $1/7 < \gamma \leq 3/7$
and separability for $0 \leq \gamma \leq 1/7$, which (for $\gamma
\rightarrow -\gamma$) recovers the result of
Ref.~\cite{horodecki99c}. The regions of separable, PPT entangled
and NPT entangled states on the boundary plane are illustrated in
Fig.~\ref{figplanesep}.

The plane $Pl_1$ representing the entanglement witness also intersects
the PPT cone inside the pyramid; thus the bound entangled states extend
from the boundary plane into the volume. The region of bound entanglement
is then confined by the surface of the PPT cone
$\alpha = ( -2 + 11\beta - \gamma + 3 \sqrt{\Delta})/16$, the plane $Pl_1$, and
the boundary plane.

\begin{figure}
  \includegraphics[width=0.7\textwidth]{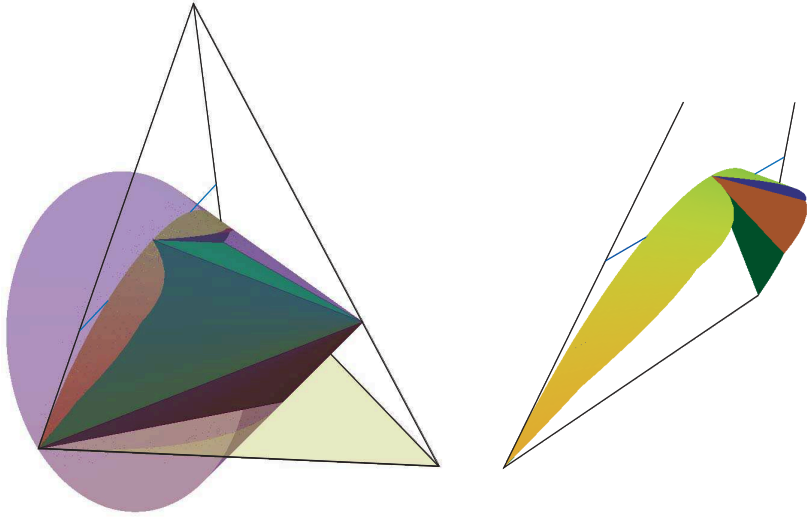}\\
  \caption{(Color online) Right: The detected regions of bound entanglement extending from the
    boundary plane (enlarged) as intersections of the planes $Pl_1$ (green), $Pl_2$
    (red) and $Pl_3$ (blue) (from bottom to top) corresponding to three entanglement witnesses.
    Left: The polygon (green) encloses the states that are necessarily
    separable. In between the polygon and the surface of the PPT cone lie the detected
    volumes of bound entanglement.} \label{figpolygon}
\end{figure}

We can detect regions of bound entanglement that reach even further into
the pyramid by using other entanglement witnesses that correspond to other
planes intersecting the cone. We construct again witnesses with our
geometric method on lines starting on the boundary plane around the
Horodecki line and such we obtain the smallest possible value of $\lambda$,
i.e., $\lambda_{\rm{min}}^{\rm{tot}} = (3 + \sqrt{13}\,)/8
\simeq 0.826$, for the starting point with $\epsilon_0 = (-25+7\sqrt{13}\,)
\,/\,2 \,\simeq\, 0.12$ and $\gamma = \sqrt{5 + 11\epsilon_0 /3
- 5\epsilon_0^2 /12}\,/\,7 \,\simeq\,0.35$. The state on this line with
$\lambda_{\rm{min}}^{\rm{tot}}$ is the outermost point of the bound entangled volume
(the green tip in Fig.~\ref{figconepyramid}) that reaches into
the PPT cone. The entanglement witness $C_\lambda$
corresponding to $\lambda_{\rm{min}}^{\rm{tot}}$ gives the plane
equation $Pl_2: \, \alpha = (-524+148 \sqrt{13}\,)^{-1} [ 16 \left(-7+2
\sqrt{13}\,\right)-4 \left(-5+\sqrt{13}\,\right) \beta +\left(-94+26
\sqrt{13}+3(-5+\sqrt{13}\,)\sqrt{2\epsilon_0}\,\right) \gamma ]$.
It offers a new boundary for the bound entangled region that extends
further into the pyramid (but detects fewer bound entangled states on the
boundary plane of the pyramid). A final extension we find by considering the
line that starts at a point $\gamma = 2/7$ of the intersection curve
of $Pl_1$ with the PPT cone. The minimal $\lambda$ such that $C_\lambda$
\eqref{clambda} represents an entanglement witness according to Lemma~\ref{lemqutrit}
is $\lambda_{\rm min} = 7 (2328 + 331 \sqrt{39}\,)/32763 \,\simeq\, 0.94$,
and the corresponding plane $Pl_3: \,\alpha = (150-18
\sqrt{39}\,)^{-1}\left[24-2 \sqrt{39}+\left(-42+9 \sqrt{39}\,\right)
\beta -6 \left(-5+\sqrt{39}\,\right) \gamma \right]$ provides a new
boundary for the bound entangled states reaching yet a bit further
into the volume of the pyramid. The planes $Pl_1, Pl_2, Pl_3$ together with
the detected volumes of bound entangled states we have depicted in Fig.~\ref{figpolygon}.\\

Naturally the question arises if we can detect all regions of NPT
entangled, PPT entangled and separable states for our
three--parameter family of states \eqref{famstates}. As a first step
to an answer we can identify a polygon of states that have to be
separable for sure by connecting all outermost separable states we
know so far in a convex manner. These known states are the boundary
of the PPT states for $\gamma = 0$, which form a trapezoid, and the
boundary of the separable states on the boundary plane.
The resulting polygon can be seen in Fig.~\ref{figpolygon}.

Numerical calculations imply that the set of separable states is
larger than the constructed polygon, and that bound entanglement is
mainly concentrated in the region we detected. These results require
numerical minimizations to obtain the needed entanglement witnesses,
and the detailed procedure will be discussed in a forthcoming
article.

Summarizing, we find new, large regions of bound entangled states in
a three--parameter family of the magic simplex of two qutrits. We
detect the definite regions of separable, PPT entangled and NPT
entangled states on a boundary plane of the family of states by
using entanglement witnesses that are constructed geometrically. The
regions of bound entangled states extend from the plane into the
pyramid of the states forming such large volumes of bound
entanglement that include the Horodecki states as a small line.
Finally, we construct a convex subset of the three--parameter
family, a polygon, that encloses states that are necessarily
separable.

We would like to thank Alexander Ableitinger for helpful comments.
This research has been financially supported by the FWF project
CoQuS No W1210-N16 of the Austrian Science Foundation and by the
F140-N Research Grant of the University of Vienna.

\bibliography{references}

\end{document}